\documentclass[11pt,twoside]{article}
\usepackage{asp2010}

\resetcounters

\begin{document}

\title{ A Provocative Summary: Is There Any Unified Model for Triggering Active Galactic Nuclei ?}

\author{ Yoshiaki Taniguchi$^1$}
\affil{$^1$Research Center for Space and Cosmic Evolution, Ehime University,
Japan, Bunkyo-cho 2-5, Matsuyama, Ehime 790-8577, Japan}

\begin{abstract}
A provocative summary of the conference on gGalaxy Mergers in 
an Evolving Universeh is presented. The conference includes 
the following issues; 1) Mergers at low redshift, 
2) Merger dynamics and numerical simulations, 
3) Mergers in galaxy evolution, 4) Mergers and active galactic 
nuclei (AGNs), 5) Mergers at high redshift and submm galaxies, 
6) Feedback from starbursts and AGNs, 6) Chemical evolution of 
galaxies, and 7) Results from new surveys. 
Although there are many important issues related to galaxy mergers 
and nuclear activities in galaxies, in this summary, l
et us focus on following the question; {\it Is there any unified 
model for triggering nuclear activities ?} Given key 
observational properties of galaxies with AGNs, a possible unified
 model is discussed as an evolutionary model from starbursts to
 AGNs based on minor and major mergers between/among galaxies.

\end{abstract}

\section{Introduction}

We observe galaxies with an active galactic nucleus (AGN) together
 with normal and starburst galaxies in the local universe 
(e.g., Ho et al. 1997). AGNs are considered to be powered by the 
gas accretion onto a supermassive black hole (SMBH) resided in a 
nucleus of galaxies (Rees 1984). In this respect, such AGN phenomena
 are basically different from the ordinary evolution of galaxies that
 are governed by various star formation history. However, it has been
 shown that the mass of SMBHs is well correlated with that of the
 spheroidal component of galaxies (bulges and elliptical galaxies)
 regardless of the presence of AGN phenomena (e.g., Magorrian et al. 
1998).  This suggests strongly the co-evolution between SMBHs 
and galaxies although we have not yet understood what physical 
processes play a key role. 

More importantly, SMBHs are ubiquitously present in the nucleus of
 galaxies and thus the AGN phenomena should be controlled by some
 triggering mechanism such as gas fueling onto the SMBH. This means
 that all AGN phenomena are transient processes during the course of
 evolution of galaxies. This is also advocated both by the evolution
 of quasar luminosity function (e.g., Boyle et al. 1992; Ikeda et al.
 2011 and references therein) and by the so-called dead quasar problem
 (Rees 1990). 

Another type of activity in nuclear regions of galaxies is starburst
 phenomena in which more than 10,000 massive stars are formed during
 a short time interval (Balzano 1983). Since massive stars could provide
 compact objects such as neutron stars and stellar-sized black holes,
 it has been also discussed that there may some possible relationships
 between starburst and AGN phenomena; i.e., starburst-AGN connections
 (e.g., Taniguchi 2003, 2004). Here we should remind that the starburst
 activity also needs some triggering mechanisms because a lot of dense
 gas must be accumulated in the nuclear region of galaxies. 
In fact, the starburst is intrinsically different from 
ordinary star formation in galactic disks (Daddi et al. 2010). 

In summary, we have to consider various aspects when we investigate
 the evolution of nuclear activities and chemical and dynamical
 properties of galaxies as a function of the cosmic age; see Fig. 1.
 In this article, however, let us focus on the importance of mergers
 of galaxies toward the understanding of triggering both starbursts
 and AGNs.

\begin{figure}[htbp]
\begin{center}
\includegraphics[width=130mm]{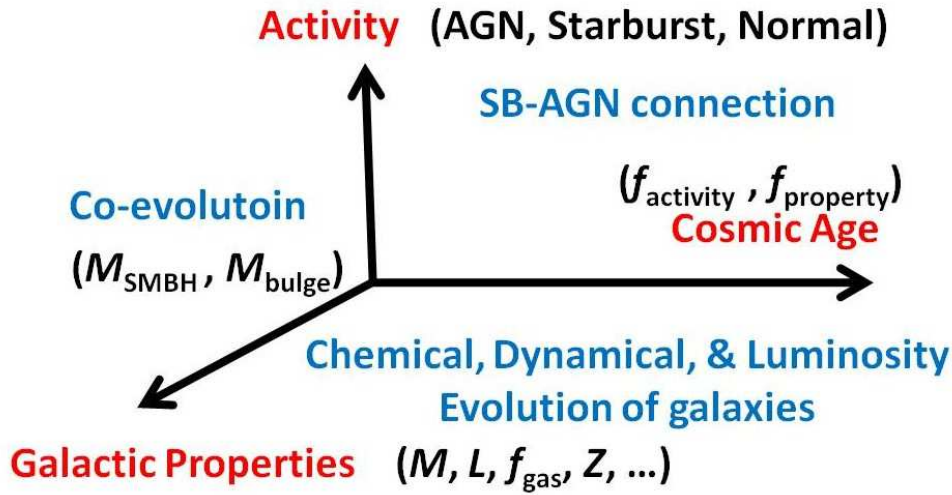}
\caption{A three-dimensional, schematic summary of evolution of both starbursts and AGNs during the course of galaxy evolution. (1) There are a number of important properties of galaxies such as the mass ($M$), luminosity ($L$), gas fraction ($f_{\rm gas}$), metallicity ($Z$), and so on. (2) There are two distinct types of nuclear activities, AGNs and starbursts, in addition to the normal phase of nuclei. (3) It is reminded that these two parameters (the nuclear activities and the galaxy properties) that are denoted as $f_{\rm activity}$ and $f_{\rm property}$, respectively, are functions of the cosmic age. If the AGN phenomena are in part related to the growth of SMBHs, the co-evolution between SMBHs and galaxies is expected.
}
\end{center}
\end{figure}

\section{Possible Unified Models for Triggering AGNs}

In this summary, we are interested in both starbursts and AGNs. 
However, for simplicity, we first consider what a mechanism can trigger
 AGN phenomena and then how a physical connection between starbursts
 and AGNs.  Therefore, our questions are summarized as follows.

\vspace{0.5cm}

\begin{itemize}

\item [1] Why do some galaxies show AGN phenomena although almost all galaxies have a SMBH in their nuclear regions ? 

\item [2] Do galaxy mergers play an essentially important role for triggering them ? 

\item [3] If the answer is yes, how do they work ? 

\end{itemize}

\vspace{0.5cm}

In order to give possible answers for these questions, let us start to adopt the following theorem. 

\begin{center}
{\bf Theorem:  There is the unified model for triggering AGNs.}
\end{center}

This means the following scenario; If an ordinary (i.e., non-active)
 galaxy has a certain triggering mechanism, it must {\it inevitably}
 evolve to a galaxy with an AGN. Namely, we postulate that there is
 the only one triggering mechanism of AGNs (see Fig. 2). 
Therefore, the question is what the only one mechanism is.

\begin{figure}[htbp]
\begin{center}
\includegraphics[width=130mm]{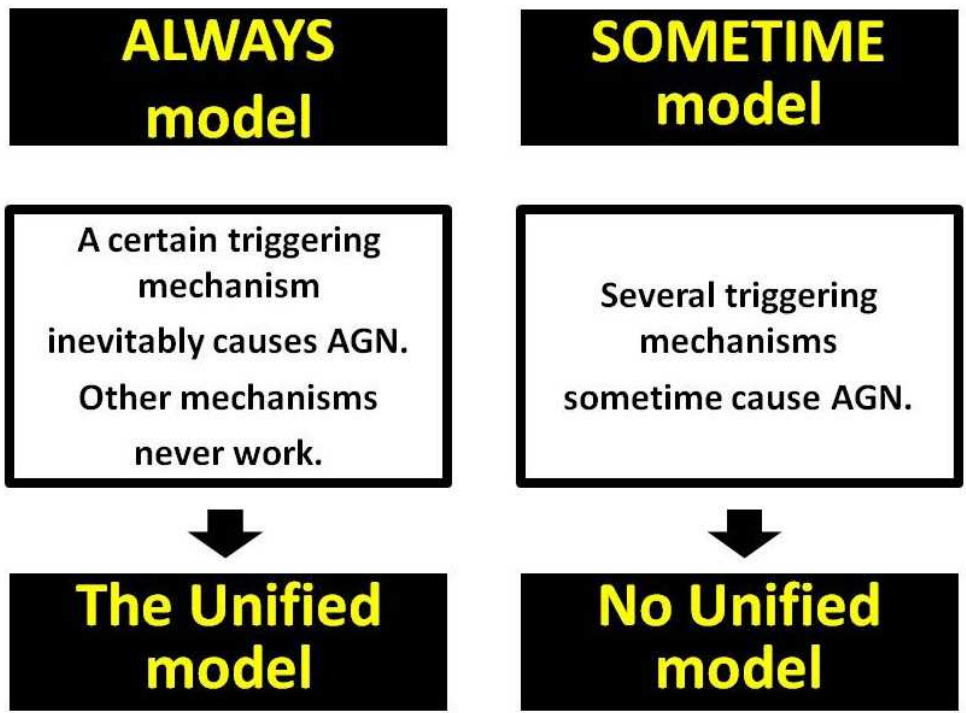}
\caption{"Always model" vs. "Sometime model". If there is an "Always Model", that is the unified model. 
}
\end{center}
\end{figure}

As possible triggering mechanisms for starbursts and AGNs, 
the following items have been discussed intensively (see, for a review,
 Taniguchi 1999);

\vspace{0.5cm}

\begin{enumerate}

\item Secular-evolution fueling in galactic disks

\item Bar-driven fueling

\item Minor-merger-driven fueling

\item Galaxy-interaction-driven fueling

\item Major-merger-driven fueling

\end{enumerate}

\vspace{0.5cm}

\subsection{Triggering Seyfert Nuclei}

First, let us consider about Seyfert nuclei in the local universe 
because many observational properties have been investigated 
to date.
At this starting point of discussion, we can omit the
major-merger-driven fueling that is considered to play 
an important role in triggering ultraluminous infrared 
galaxies (ULIRGs) and quasars (Sanders et al. 1988). 
We shall come back to this mechanism later.

First, we can exclude the secular-evolution fueling. 
Since, if this is the case, we cannot specify when and how 
the Seyfert activity could be triggered, this idea does not fit 
to any unified triggering mechanism.

As described in Taniguchi (1999), both the bar-driven
and interaction-driven fueling mechanisms also do not fit.
The fraction of barred spiral galaxies is approximately 70 percent 
for normal and Seyfert galaxies. In other word, Seyfert galaxies 
do not show preference of barred spirals. As for the galaxy interaction,
 only 10 percent of Seyfert galaxies have their companion galaxies. 

Accordingly, the minor-merger-driven fueling can remain as a 
possible unified triggering mechanism for Seyfert activity.
Minor mergers have been often considered as a plausible 
triggering mechanism for nuclear starbursts (e.g., Mihos \& 
Hernquist 1994; Taniguchi \& Wada 1996). 
In particular, minor mergers with a nucleated satellite are 
expected to surely provide gas fueling very close toward a SMBH 
in nuclei of galaxies. Since their tidal disturbance is 
usually weak, it is often difficult to find evidence for
 minor mergers. However, the observed excess of ringed structures 
in Seyfert disks, and S0 or amorphous morphology of Seyfert hosts 
suggest past minor merger events in them. Another line of evidence
 for minor mergers is advocated by randomly-oriented narrow-line
 regions with respect to the host disks in Seyfert galaxies. 
Such properties are naturally explained by an idea that 
the orbital plane of a dusty torus should be the dynamically 
preferred plane governed by the orbital motion of a merging 
satellite galaxy. Although these are not convincing pieces of 
evidence, the minor-merger-driven fueling can be appreciated 
as a plausible unified mechanism for triggering AGN (Taniguchi 1999). 

\subsection{Triggering Quasars}

Quasars are the most luminous class of AGNs whose luminosity 
exceeds $10^{12} L_{\odot}$.
Although their origin has 
been mysterious for a long time, the discovery of ULIRGs led to 
the following formation mechanism; major mergers between two 
gas-rich galaxies could result in the formation of quasars through 
the phase of ultra luminous starbursts
(Sanders et al. 1988). This idea is also reinforced by 
disturbed morphology of quasar host galaxies (Bahcall et al. 1995).
 Although some quasar hosts show undisturbed morphology like elliptical
 galaxies, they can also be interpreted as advanced mergers. 

Important physical processes involved in the major-merger scenario 
are considered to be the growth of SMBHs. The mass of SMBHs in 
typical disk galaxies may range from 1 million to 10 million $M_{\odot}$.
However, a typical mass of SMBHs in quasar nuclei ranges from 100 
million to 1 billion $M_{\odot}$. Therefore, the growth of SMBHs 
could be helped by supply of a large number of compact objects as 
remnants of massive stars formed in ultra luminous 
starbursts (Taniguchi, Ikeuchi, \& Shioya 1999). 

Another concern is how many galaxies merged into one to make 
a quasar. At least, two gas-rich giant galaxies are necessary to
 possess a lot of cold molecular gas of $\sim 10^{10} M_{\odot}$ 
(Sanders et al. 1988). In the local universe, there are a number 
of compact galaxies whose merging time scale is of order of 1 
billion years (Hickson 1982). They are also likely precursors of 
quasars via ULIRGs (Taniguchi \& Shioya 1988). 

In summary, one has the following scenario for the formation of quasars;

\vspace{0.5cm}

\begin{itemize}

\item [{\bf Step 1}.] {A major merger between or among gas-rich galaxies.}

\item [{\bf Step 2}.] {The formation of a ULIRG that helps the growth of a SMBH in the merger center.}

\item [{\bf Step 3}.] {The formation of quasar after the ULIRG phase.}

\end{itemize}

\vspace{0.5cm}

It should be noted that the major merger is the most important 
triggering mechanism for the quasar formation. In other word, 
we need any other triggering mechanisms here, providing the 
unified formation mechanism of quasars.

\section{Possible Unified Models for Triggering AGNs: An Evolutionary Unified Model for the Starburst-AGN Connection}

We have argued whether possible unified models can be available 
for the understanding triggering both Seyferts and quasars in 
the two previous sections. We have found that the 
minor-merger-driven fueling provides only one mechanism for 
triggering Seyferts while the major-merger one does that for 
triggering quasars. In other word, one can postulate that only 
mergers can trigger the AGN activity regardless of their AGN luminosity.

Here we remind for the case of quasars that the ultra luminous 
starburst phase comes first and then the quasar activity turns on 
in the final phase of merger evolution. We interpret this as that 
such starbursts are necessary to grow up a SMBH in a merger 
center (Taniguchi, Ikeuchi, \& Shioya 1999). If this is also 
the case in triggering Seyferts, it is suggested that the 
so-called nuclear starburst phase comes first and then the 
Seyfert activity turns on. Since it is known that such nuclear 
starbursts are induced by minor mergers 
(e.g., Mihos \& Hernquist 1994; Taniguchi \& Wada 1996),
 the above suggestion seems to make sense. 

Accordingly, one can have a merger-driven, evolutionary unified 
model from the starburst to the AGN. This model is illustrated 
in Fig. 3 (see also Taniguchi, Wada, \& Murayama 1996).

\begin{figure}[htbp]
\begin{center}
\includegraphics[width=100mm]{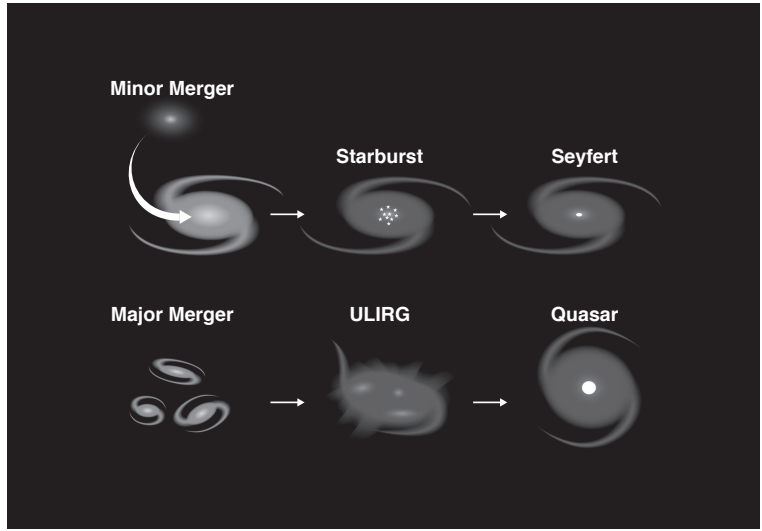}
\caption{The merger-driven unified model for triggering AGNs. This figure is made by Ryoichi Saito.
}
\end{center}
\end{figure}

\section{Summary and Future Prospects}

Our conclusions are summarized below;

\begin{enumerate}

\item Secular-evolution fueling in galactic disks {\bf never works}.

\item Bar-driven fueling {\bf never works}.

\item Minor-merger-driven fueling {\bf works} for triggering Seyferts.

\item Galaxy-interaction-driven fueling {\bf never works}.

\item Major-merger-driven fueling {\bf works} for triggering quasars.

\end{enumerate}

One may feel that the above conclusions are somewhat 
drastic ones. Actually, non-axisymmetric structure such as bars 
(e.g., Shlosman et al. 1989; Wada \& Habe 1995) and 
non-axisymmetric perturbation such as galaxy interactions 
(e.g., Noguchi 1988) have been often considered as viable gas 
fueling mechanisms in AGNs.  However, one serious physical 
difficulty in such fueling mechanisms is that it is generally 
difficult to reduce the angular momentum of disk gas 
($\sim$ 1 - 10 kpc scale) toward the inner $\sim$ 0.01 or 0.001 pc 
around a SMBH (e.g., Peterson 1997). This is the main reason 
why we discard both the secular-evolution and the bar-driven 
fueling mechanisms in the proposed unified model.

In order to avoid the angular momentum problem, we would like to 
note that minor mergers with a nucleated satellite are more 
preferable because the satellite nucleus surely can reach to 
the center of the host galaxy (Taniguchi \& Wada 1996). 
Note also that Sanders et al.'s scenario for the quasar 
formation postulates major mergers between tow nucleated galaxies.

In summary, the unified model for triggering both starbursts and 
AGNs means that any nuclear activities are due not to nature 
but to nurture. In other words, if a galaxy were isolated during 
the course of its evolution, it did not have a SMBH and thus never
 became to an AGN-hosting galaxy. Nuclear starbursts, too. 
All nuclear activities are driven by second events 
(minor and major mergers) in terms of the so-called 
hierarchical clustering model of our universe.

\acknowledgements

We would like to thank all the LOC members of this conference, 
in particular, Paul Ho, Wei Hsin Sun, and Wei Hao Wang.  
We would also like to thank Nick Scoville, Dave Sanders, Dave Koo, 
and many friends at the conference for their stimulating discussion.
 Special thanks are due to Ryoichi Saito for his wonderful drawing 
of Fig. 3. This work has been supported in part by JSPS grants 
(Nos. 23244031 \& 23654068).


\begin{references}
\reference{}{Bahcall, J. N., et al. 1997, ApJ, 479, 642}
\reference{}{Balzano, V. A. 1983, ApJ, 268, 602}
\reference{}{Boyle, B J., et al. 1990, MNRAS, 317, 1014}
\reference{}{Daddi, E., et al. 2010, ApJ, 714, L118}
\reference{}{Hickson, P. 1982, ApJ, 255, 382}
\reference{}{Ho, L. C., Filippenko, A. V., \& Sargent, W. L. W. 1997, ApJ, 487, 591}
\reference{}{Ikeda, H., et al. 2011, ApJ, 728, L25}
\reference{}{Magorrian, J., et al. 1998, AJ, 115, 2285}
\reference{}{Mihos, C., \& Hernquist, L. 1994, ApJ, 425, L13}
\reference{}{Noguchi, M. 1988, A\&A, 203, 259}
\reference{}{Peterson, B. M. 1997 An Introduction to Active Galactic Nuclei (Cambridge University Press)}
\reference{}{Rees, M. J. 1984, ARA\&A, 22, 471}
\reference{}{Rees, M. J. 1990, Science, 247, 817}
\reference{}{Sanders, D. B., et al. 1988, ApJ, 325, 74}
\reference{}{Shlosman, I., Frank, J., \& Begelman, M. C. 1989, Nature, 338, 45}
\reference{}{Taniguchi, Y. 1999, ApJ, 524, 65}
\reference{}{Taniguchi, Y. 2003, ASPC, 289, 353}
\reference{}{Taniguchi, Y. 2004, Progress of Theoretical Physics Suppl., 155, 20}
\reference{}{Taniguchi, Y., Ikeuchi, S., \& Shioya, Y. 1999, ApJ, 514, L9}
\reference{}{Taniguchi, Y.,\& Shioya, Y. 1998, ApJ, 501, L167}
\reference{}{Taniguchi, Y., \& Wada, K. 1996, ApJ, 469, 581}
\reference{}{Taniguchi, Y., Wada, K., \& Murayama, T. 1996, Revista Mexicana de Astronomia y Astrofisica Serie de Conferencias, Vol. 6, 1st Guillermo Haro Conference on Astrophysics: Starburst Activity in Galaxies, Puebla, Pue., Mexico, April 29-May 3, 1996, p. 240}
\reference{}{Taniguchi, Y., et al. 2012, ApJ, submitted}
\reference{}{Wada, K., \& Habe, A. 1995, MNRAS, 277, 433}
\end{references}
\end{document}